\documentclass[english,pra,reprint,superscriptaddress]{revtex4-1}
\usepackage{letltxmacro}
\LetLtxMacro{\oldauthor}{\author}
\global\long\def\author[#1]#2{\oldauthor{#2}}
\global\long\def\affil[#1]#2{\affiliation{#2}}

\usepackage{amsmath, mathtools}
\usepackage{amssymb}
\usepackage{wasysym}
\usepackage{graphicx}
\usepackage{import}
\usepackage{color}
\usepackage{transparent}
\usepackage{braket}
\usepackage{overpic}
\usepackage{leftidx}
\usepackage[bookmarks=true, breaklinks=false, pdfborder={0 0 0},backref=false, colorlinks=false]{hyperref}

\begin{document}


\title{Quantum Gates for Propagating Microwave Photons}
\author[1]{Roope Kokkoniemi \thanks{roope.kokkoniemi@aalto.fi}}
\affiliation{QCD Labs, COMP Centre of Excellence, Department of Applied Physics, Aalto University, FI-00076 Aalto, Finland}
\author[1]{Tuomas Ollikainen}
\affiliation{QCD Labs, COMP Centre of Excellence, Department of Applied Physics, Aalto University, FI-00076 Aalto, Finland}
\author[1,2]{Russell E. Lake}
\affiliation{QCD Labs, COMP Centre of Excellence, Department of Applied Physics, Aalto University, FI-00076 Aalto, Finland}
\affiliation{National Institute of Standards and Technology, Boulder, Colorado 80305, USA}
\author[1]{Sakari Saarenp\"a\"a}
\affiliation{QCD Labs, COMP Centre of Excellence, Department of Applied Physics, Aalto University, FI-00076 Aalto, Finland}
\author[1]{Kuan Yen Tan}
\affiliation{QCD Labs, COMP Centre of Excellence, Department of Applied Physics, Aalto University, FI-00076 Aalto, Finland}
\author[1]{Janne I. Kokkala}
\affiliation{QCD Labs, COMP Centre of Excellence, Department of Applied Physics, Aalto University, FI-00076 Aalto, Finland}
\author[1,3]{Ceren B. Da\u{g}}
\affiliation{QCD Labs, COMP Centre of Excellence, Department of Applied Physics, Aalto University, FI-00076 Aalto, Finland}
\affiliation{Physics Department, University of Michigan, 450 Church St., Ann Arbor, MI 48109-1040, USA}
\author[1]{Joonas Govenius}
\affiliation{QCD Labs, COMP Centre of Excellence, Department of Applied Physics, Aalto University, FI-00076 Aalto, Finland}
\makeatletter
\author[1]{Mikko M\"{o}tt\"{o}nen}
\affiliation{QCD Labs, COMP Centre of Excellence, Department of Applied Physics, Aalto University, FI-00076 Aalto, Finland}
\makeatother

\begin{abstract}
We report a generic scheme to implement transmission-type quantum gates for propagating microwave photons, based on a sequence of lumped-element components on transmission lines. By choosing three equidistant superconducting quantum interference devices (SQUIDs) as the components on a single transmission line, we experimentally implement a magnetic-flux-tunable phase shifter and demonstrate that it produces a broad range of phase shifts and full transmission within the experimental uncertainty. Together with previously demonstrated beam splitters, these phase shifters can be utilized to implement arbitrary single-qubit gates. Furthermore, we theoretically show that replacing the SQUIDs by superconducting qubits, the phase shifter can be made strongly nonlinear, thus introducing deterministic photon--photon interactions. These results critically complement the previous demonstrations of on-demand single-photon sources and detectors, and hence pave the way for an all-microwave quantum computer based on propagating photons.
\end{abstract}

\maketitle

{\it Introduction.}---Since its initial theoretical considerations in the 1980s~\cite{Feynman1982}, quantum computing has been an active area of research thanks to its envisioned superior performance in certain computational problems~\cite{simon1997power,childs2010quantum}. For example, detailed simulations of many-particle quantum systems are out of reach of classical computers, but ideally suit controllable quantum systems~\cite{houck2012chip}. However, the realization of a high-fidelity qubit register~\cite{kelly2015state} remains a major challenge in the implementation of a large-scale quantum computer. Among many different proposals~\cite{ladd2010quantum}, photonic systems constitute an interesting candidate for the register since they exhibit weak decoherence~\cite{o2007optical} and photons can be directly used for fast and secure communication~\cite{BB84}.

The minimal requirements for the realization of a path-encoded photonic quantum computer are the following~\cite{RevModPhys.79.135}: high-fidelity on-demand photon sources and single-photon detectors are needed for the initialization and measurement of the quantum states, respectively. In addition, tunable single-qubit gates are required to program desired unitary evolutions. A convenient way to realize arbitrary single-qubit gates is to use static beam splitters and tunable phase shifters. Finally, one needs to introduce nonlinearity for qubit--qubit interactions in order to create entangling two-qubit gates. In principle, the nonlinearity brought about from the measurements and classical feedback is sufficient~\cite{knill2001scheme} but nondeterministic, and hence, in practice, it requires a large overhead in the other resources. Significant reduction of the overhead is provided by a paradigm referred to as one-way quantum computing~\cite{raussendorf2001one} where a so-called cluster state including all desired correlations is built and measured. Even less overhead is needed if two-qubit gates are directly and deterministically imposed on the register, but high-fidelity implementation of such gates is challenging.

Thanks to the long and successful history in the optics research and industry, photons at optical wavelengths have attracted the most attention as candidates for photonic qubits~\cite{o2009photonic}.
Here, on-demand single-photon sources have reached 65\% efficiencies~\cite{somaschi2016near,ding2016demand} and above 90\% system detection efficiency has been reported~\cite{Lita:08,marsili2013detecting}. Traditionally, tunable optical phase shifters have been slow to operate~\cite{Humphreys2014,Matthews2009,Smith:09}, but recent electro-optical phase shifters~\cite{PhysRevLett.108.053601} have potential for 100-GHz operation~\cite{kanno2010120}. However, since their operation principle is based on the rather weak Pockels effect, large bias voltages or relatively long electrodes are needed in order to achieve reasonable phase shifts~\cite{wooten2000review,wang2014gallium}, hindering the scalability of the optical quantum computer. Furthermore, optical photons interact relatively weakly with nonlinear matter, rendering it very challenging to implement deterministic two-qubit gates~\cite{turchette1995measurement}, and hence adding the above-discussed overhead related to measurement-based gates.

Recent years have witnessed great progress in the implementation of high-fidelity superconducting qubits operating at microwave frequencies~\cite{blais2004cavity,kelly2015state, devoret2013superconducting,dicarlo2010preparation,wallraff2005approaching,chiorescu2003coherent}. These nonlinear circuit elements can be engineered to interact very strongly~\cite{yoshihara2016superconducting}
with single photons~\cite{bozyigit2011antibunching,niemczyk2010circuit},
providing opportunities for introducing deterministic photon--photon interactions \cite{koshino2016tunable}. In practice, on-demand single-photon sources in the microwave regime have already reached above 80\% efficiencies~\cite{houck2007generating,peng2016tuneable,george2016multiplexing}, thus surpassing the optical sources.
Qubit-based single-photon microwave detectors~\cite{romero2009, PhysRevLett.102.173602,PhysRevLett.107.217401,PhysRevA.84.063834,PhysRevLett.110.053601,PhysRevLett.111.053601,PhysRevLett.112.093601,PhysRevB.90.035132,PhysRevA.91.043805,narla2016robust} have recently reached a quantum efficiency of 66\%~\cite{inomata2016single} and calorimetric detectors~\cite{ullom2015review,pekola2013calorimetric}
have taken a leap towards the single-photon regime~\cite{govenius2016detection}. Owing to fully compatible fabrication techniques, all these components can be conveniently integrated on the same chip~\cite{bozyigit2011antibunching},
thus rendering propagating microwave photons
an attractive alternative to optical photons in realizing a photonic quantum computer.

However, no quickly tunable, compact, and high-fidelity phase shifter for propagating microwave photons has been demonstrated to date. A transmission line with a current-tunable kinetic inductance can be used as a phase shifter~\cite{anlage1989current,eom2012wideband}, but this has a major drawback since changing the inductance of the transmission line also changes its characteristic impedance. Hence, a tunable impedance-matching circuitry would be necessary for connecting this kind of a phase shifter to the rest of the ultralow-loss circuitry.

\begin{centering}
\begin{figure}
\includegraphics{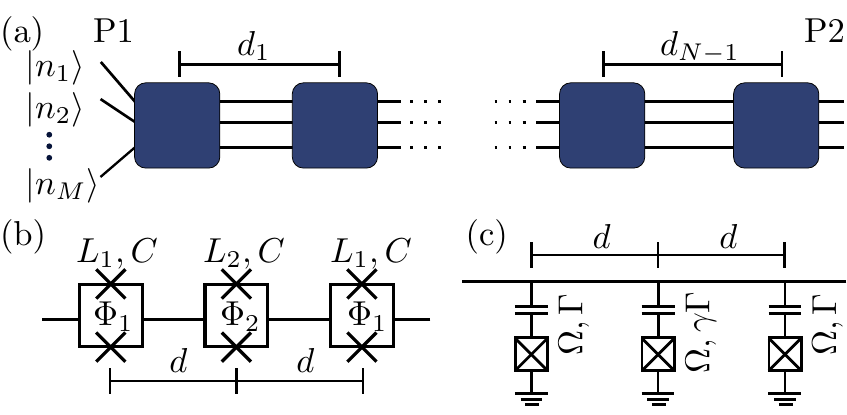} 
\caption{(a) Generic scheme for the implementation of multi-qubit quantum gates for propagating microwave photons. Each gate consists of $N$ elements, depicted by the blue boxes, separated by the distances $\{d_k\}$, where $k\in\left\{1,2,...,N-1\right\}$, and connected to $M$ transmission lines. In the dual-rail representation, the basis states of the quantum register compose of the Fock states  $\ket{n_1,n_2,\dots,n_M}$, where $\left(n_{2m},n_{2m-1}\right)\in\left\{(1,0),(0,1)\right\}, m\in\left\{1,2,...,M/2\right\}$. The input port of the gate is denoted by P1 and the output port by P2. (b,c) Circuit diagram of a phase shift gate where the elements are (b) SQUIDs or (c) capacitively coupled superconducting qubits, separated by distance $d$. The definitions of the other symbols are given in the main text.}
\label{fig:lumped_element_model}
\end{figure}
\end{centering}

In this Letter, we propose a generic scheme to implement quantum gates for propagating photons based on a sequence of components acting on transmission lines as shown in Fig.~\ref{fig:lumped_element_model}(a). The properties and distances between the components are chosen to realize full transmission with the desired effect on the photons. In this framework, we focus on two concrete and useful examples: a linear and a nonlinear phase shifter. Namely, we demonstrate a compact linear phase shifter based on three equidistant superconducting quantum interference devices (SQUIDs) interrupting a transmission line, see Fig.~\ref{fig:lumped_element_model}(b). The SQUIDs can be tuned in situ at timescales set by their plasma frequency which is typically of the order of 10 GHz. Furthermore, we theoretically show that a strongly nonlinear phase shifter can be implemented if the SQUIDs are replaced by superconducting qubits capacitively coupled to the transmission line, see Fig.~\ref{fig:lumped_element_model}(c).

{\it Linear phase shifter.}---Let us derive an analytical expression for the transmission coefficient of the SQUID-based phase shifter shown in Fig.~\ref{fig:lumped_element_model}(b) employing the quantum network theory~\cite{Yurke1984}. We model the SQUIDs as linear $LC$ oscillators, which is justified by the fact that the photon current is well below the typical critical currents of the SQUIDs. The Josephson inductances of the SQUIDs are given by
\begin{equation}\label{eq:ind}
L_i=\frac{\Phi_{0}}{4\pi I_{\mathrm{c},i}\left|\cos\left(\pi\frac{\Phi_{i}}{\Phi_{0}}\right)\right|},
\end{equation}
where $\Phi_{0}$ is the magnetic flux quantum and $I_{\mathrm{c},i}$ and $\Phi_i$ are the critical current and the magnetic flux threading the $i$th SQUID. The parallel capacitance, $C$, in the model arises from the junction capacitance and possible stray capacitances between the different sides of the SQUID. For simplicity, we have assumed that all Josephson junctions are identical and that the geometric inductance is negligible in Eq.~\eqref{eq:ind}.

We tune the inductances of the SQUIDs at the left and right ends of the chain to be equal, $L_1$. The middle SQUID inductance is denoted by $L_2$. Dissipation is negligible in the short superconducting waveguides we consider, and hence we do not account for it in our model. After a straightforward calculation~\cite{SI}, we find that the forward transmission coefficient of the voltage through the SQUID chain is given by $S_{21}^\text{S}(\omega)=f^\text{S}(\omega)/g^\text{S}(\omega)$, where the functions $f^\text{S}$ and $g^\text{S}$ assume the forms
\begin{equation} \label{eq:a_r_factor}
f^\text{S}(\omega)=8Z_{0}^{3}e^{2i\varphi}\left(CL_{2}\omega^{2}-1\right)\left(CL_{1}\omega^{2}-1\right)^{2},
\end{equation}
and
\begin{align} \label{eq:squit_trans_coeff_denom}
g^\text{S}(\omega)=& \left\{4Z_{0}^{2}\left(CL_{2}\omega^{2}-1\right)\left(CL_{1}\omega^{2}-1\right) +L_{2}L_{1}\omega^{2} \right.\nonumber\\
&\left.\bigr.\times\left(e^{2i\varphi}-1\right)+2i\omega Z_{0}\left[CL_{2}L_{1}\omega^{2}\left(2+e^{2i\varphi}\right) \right.\right. \nonumber \\
&\bigl.\bigl. -L_{2}-L_{1}\left(1+e^{2i\varphi}\right)\bigr]\bigr\}\left[2Z_{0}\left(CL_{1}\omega^{2}-1\right) \right. \nonumber \\
&\left. -iL_{1}\omega\left(e^{2i\varphi}-1\right)\right].
\end{align}
Here, $Z_0$ is the characteristic impedance of the transmission line, $\omega$ is the frequency of the incoming photons, $\varphi=\omega d/v$ is the phase shift due to a uniform transmission line of length $d$, and $v$ is the speed of the wave in the transmission line.

If we express the inductances $L_1$ and $L_2$ using an arbitrary real variable $\theta$ as
\begin{align}\label{eq:L1L2}
L_{1}&=\frac{2Z_{0}\sin\left(\frac{\theta}{2}\right)}{\omega\left[2C\omega Z_{0}\sin\left(\frac{\theta}{2}\right)-\cos\left(\frac{\theta}{2}+2\varphi\right)+\cos\left(\frac{\theta}{2}\right)\right]}, \\
L_{2}&=\frac{4Z_{0}\sin\left(\frac{\theta}{2}\right)\cos\left(\frac{\theta}{2}+2\varphi\right)}{\omega\left\{ 2C\omega Z_{0}\left[\sin(\theta+2\varphi)-\sin(2\varphi)\right]+\cos(2\varphi)-1\right\} }, \nonumber
\end{align}
we find that the magnitude of the transmission coefficient is indeed unity and its phase assumes the value $2\varphi+\theta$. Thus the SQUIDs introduce a phase shift $\theta$ which is in-situ controllable by tuning the inductances according to Eqs.~\eqref{eq:L1L2}. In practice, the achievable range of phase shifts is determined by the available values for the inductances and the employed photon frequency. However, this limitation may be overcome by using more than three SQUIDs in the phase shifter. 

\begin{figure}[!h]
\def\svgwidth{8.6cm}
\includegraphics{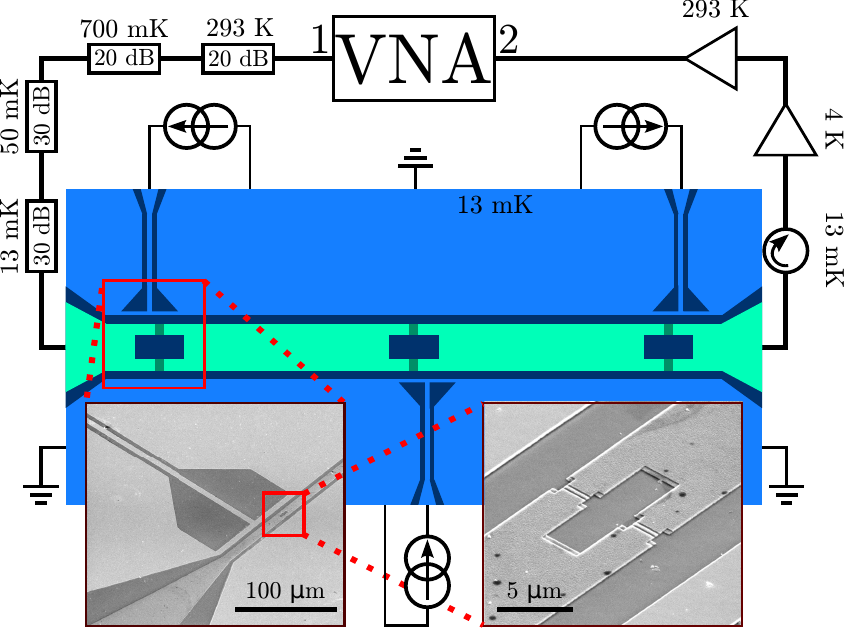}
\caption{Schematic illustration of the experimental sample and the measurement setup. The measurement signal is guided from the VNA Port 1 to the sample through attenuators at four different temperatures. The output signal from the sample is amplified with two amplifiers, one inside the cryostat and one outside, before arriving at the VNA Port 2. The sample is protected from the amplifier noise with an isolator. The three current sources allow us to control the flux bias of each SQUID individually. Here, niobium is denoted with light blue, aluminum with light green, Josephson junctions with dark green, and the substrate with dark blue color. The features are not to scale. We also show  scanning electron microscope images of the flux bias scheme (left inset) and of the first SQUID (right inset).}
\label{fig:experimental_sample}
\end{figure}

Schematic illustration of our experimental sample realizing the three-SQUID phase shifter is shown in Fig.~\ref{fig:experimental_sample}. The bonding pads for the center conductor of the waveguide are located near the left and right edges of the chip. The Al/AlO$_\textrm{x}$/Al tunnel junctions for the SQUIDs are evaporated simultaneously with the center conductor to guarantee galvanic connection.
Three broadband superconducting transmission lines provide tunable flux biases for each of the SQUID. To compensate for the cross-coupling between each flux line and the distant SQUIDs, we extract and invert the full inductance matrix~\cite{SI}. This allows us to independently control each SQUID flux.

The sample is cooled down in a commercial cryostat with a base temperature of 13 mK and measured according to the scheme presented in Fig.~\ref{fig:experimental_sample}.
We measure the transmission coefficient with a vector network analyzer (VNA). The coaxial cables connecting the output of the VNA (Port 1) to the sample are so heavily attenuated that we operate in the single-photon regime. Isolators at the output of the sample protect it from the noise of following amplifiers.

We normalize the measured transmission coefficient by a reference signal, for which that all three SQUIDs are biased at an integer multiple of a flux quantum. This reference point is chosen since the SQUID inductances are minimized here, and hence they have a minimal effect on the transmitted signal.
Namely, the critical current of each SQUID is of the order 1~$\mu$A, resulting in an impedance $i\omega L_k\approx i\times 7$~$\Omega$ at frequencies of interest. Hence, the estimated transmission amplitude for the circuit at the reference point is sufficiently close to unity, $|S_{21}^\text{S}|\approx 0.98$.

\begin{figure}[!h]
\includegraphics[width=8.6cm]{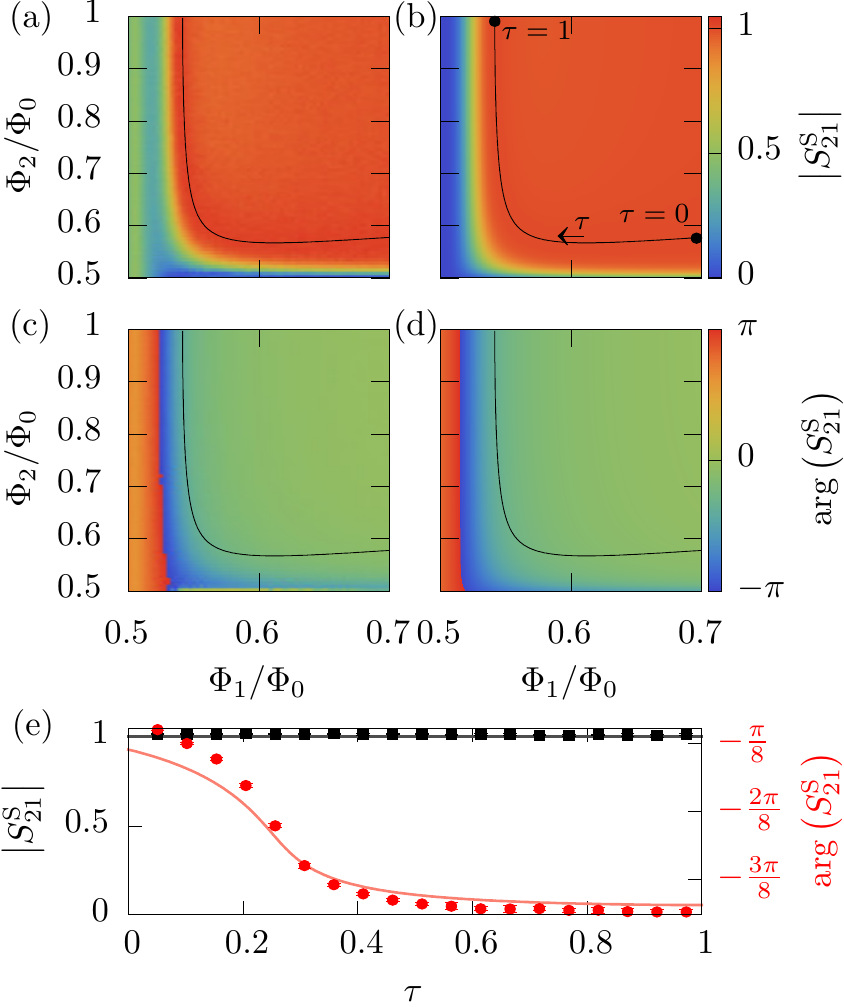}
\caption{(a--d) Normalized scattering parameter of the SQUID-based linear phase shifter, 	$S_{21}^\text{S}$, as a function of the flux through the side SQUIDs, $\Phi_1$, and through the middle SQUID, $\Phi_\mathrm{2}$, at 6.3-GHz signal frequency.
We show the measured (a) magnitude and (c) phase of the scattering parameter together with the corresponding theoretical predictions in panels (b) and (d), respectively.
We denote the relative length along the theoretical full transmission (solid line) by $\tau$ which vanishes at $\left(\Phi_1=0.7,\Phi_\mathrm{2} \approx 0.58\right)$ and is unity at $\left(\Phi_1 \approx 0.54,\Phi_2=1\right)$ as indicated in panel (b). (e) The amplitude (black color) and phase (red color) of the measured (markers) and theoretically obtained (solid lines) scattering parameters along the full-transmission curve. The measured values are obtained by fitting a quasilinear plane to the points along the curve and their eight nearest neighbors. The error bars denote the root-mean-square deviations from the fit. For the theoretical calculations, we set the critical current of the middle SQUID to $I_{\textrm{c},2}=2.2$~$\mu$A, that of the left and right SQUIDs to $I_{\textrm{c},1}=0.7$~$\mu$A,
and the SQUID capacitance to $C=26$~fF.}
\label{fig:SQUID_theory_comparison}
\end{figure}

The normalized magnitude and the phase of the measured transmission coefficient at 6.3-GHz frequency together with the corresponding theoretical predictions~\cite{SI} are shown in Figs.~\ref{fig:SQUID_theory_comparison}(a--d).
We observe nearly full transmission over a wide range of flux values. Note however that full transmission is obtained in the theoretical calculations only along the solid line in Fig.~\ref{fig:SQUID_theory_comparison}(b). Thus we show in Fig.~\ref{fig:SQUID_theory_comparison}(e) the normalized magnitude and the phase of the transmission coefficient along this line. We clearly achieve a tunable phase shift with essentially unit transmission. The small discrepancy between the observed and the theoretical phase shifts is attributed to slight deviations from the theoretical assumption that the critical currents of the side SQUIDs are equal.

{\it Nonlinear phase shifter.}---Let us theoretically consider a phase shifter with three capacitively coupled superconducting qubits as shown in Fig.~\ref{fig:lumped_element_model}(c). We analyze the behavior of this system employing a technique presented in Refs.~\cite{zheng2013persistent}, \cite{fang2014one}, and \cite{fang2015waveguide} to calculate corresponding one- and two-photon wavefunctions. The system is modeled with the usual Jaynes--Cummings
Hamiltonian employing the rotating-wave approximation. In addition, the photon wavepacket is assumed to contain frequencies in a narrow bandwidth around its central frequency, and the qubits are assumed to be accurately described by two-level quantum systems. With these approximations the full Hamiltonian reads
\begin{align}
\hat{\mathcal{H}} &= \hat{\mathcal{H}}_{\textrm{TL}} + \sum_{j=0}^{2}\left(\hat{\mathcal{H}}_{\textrm{q},j} + \hat{\mathcal{H}}_{\textrm{c},j}\right),\label{eq:finalhamiltonian}
\end{align}
where the Hamiltonians of the transmission line, $\hat{\mathcal{H}}_{\textrm{TL}}$, of $j$th qubit, $\hat{\mathcal{H}}_{\textrm{q},j}$, and of the coupling between the $j$th qubit and the waveguide mode, $\hat{\mathcal{H}}_{\mathrm{c},j}$, are given by
\begin{align}\nonumber
\hat{\mathcal{H}}_{\textrm{TL}} &= -iv\hbar\int_{-\infty}^\infty \mathrm{d}x \left[ \hat{a}_{\textrm{R}}^\dagger(x)\frac{\mathrm{d}}{\mathrm{d}x}\hat{a}_{\textrm{R}}(x) - \hat{a}_{\textrm{L}}^\dagger(x)\frac{\mathrm{d}}{\mathrm{d}x}\hat{a}_{\textrm{L}}(x) \right],\\ \nonumber
\hat{\mathcal{H}}_{\textrm{q},j} &=\hbar \Omega\hat{\sigma}_j^+\hat{\sigma}_j^-,\\ \nonumber
\hat{\mathcal{H}}_{\textrm{c},j}&=\hbar \sqrt{\frac{1}{2}\Gamma_jv}\left[ \hat{a}_{\textrm{R}}^\dagger(x_j)\hat{\sigma}_j^- + \hat{a}_{\textrm{R}}(x_j)\hat{\sigma}_j^+ \nonumber \right.\\ \nonumber
&\qquad\qquad\left.+ \hat{a}_{\textrm{L}}^\dagger(x_j)\hat{\sigma}_j^- + \hat{a}_{\textrm{L}}(x_j)\hat{\sigma}_j^+ \right].
\end{align}
Above, the operators $\hat{a}_{\textrm{L/R}}(x)$ annihilate an excitation of the waveguide mode propagating left (L) or right (R) at position $x$. The operators $\hat{a}_{\textrm{L/R}}^\dagger(x)$ create the corresponding excitations. The $j$th qubit excitation is created and annihilated by the operators $\hat{\sigma}_j^+$ and $\hat{\sigma}_j^-$, respectively. We fix the decay rate of the leftmost and rightmost qubits to the waveguide continuum to be $\Gamma_0=\Gamma_2=\Gamma$, and for the middle qubit $\Gamma_1=\gamma\Gamma$, where the coupling ratio $\gamma$ is a dimensionless parameter. Furthermore, we assume that all three qubits have an equal transition frequency $\Omega$ and set the qubits to be at locations $x_0 = -\frac{\pi}{2k}$, $x_1=0$, and $x_2 = \frac{\pi}{2k}$, where $k=\omega/v$ is the wavenumber of the photon.

\begin{figure}
\includegraphics[width=8.6cm]{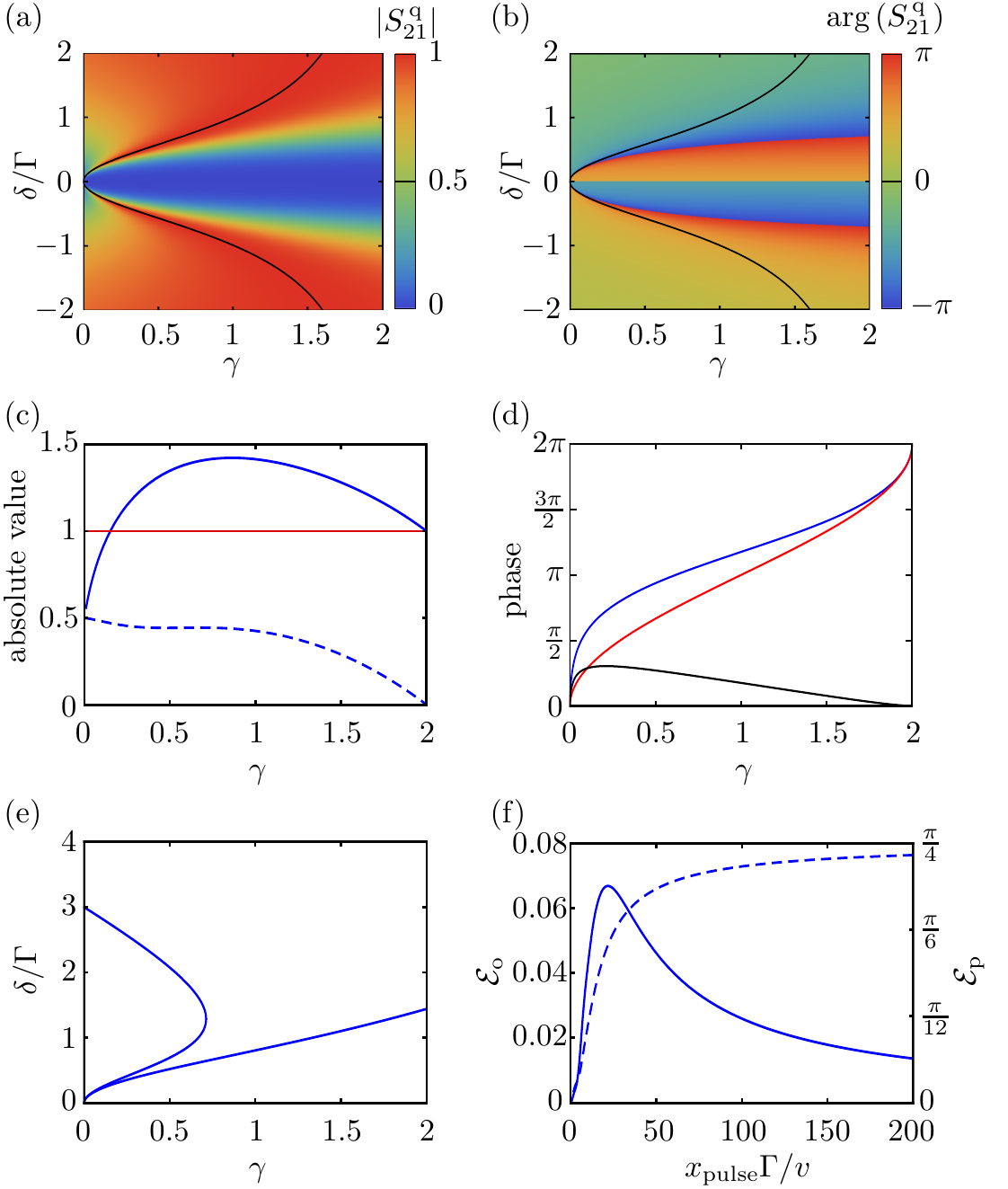}
\caption{(a) Magnitude and (b) phase of the single-photon transmission coefficient for the qubit-based phase shifter as functions of the coupling ratio and detuning. The full single-photon transmission $\delta_\text{f}$ is denoted by the solid line. The (c) magnitude and (d) phase of the transmission density for two collimated photons, $S_{21}^\text{\,q}(k_\text{f})S_{21}^\text{\,q}(k_\text{f})+B(k_\mathrm{f},0)$, (blue line) as functions of the coupling ratio. The considered wavenumber, $k_\mathrm{f}$, corresponds to full linear transmission which is shown for reference as $S_{21}^\text{\,q}(k_\text{f})S_{21}^\text{\,q}(k_\text{f})$ (red line). The difference between the nonlinear and linear phase shifts is represented by the black line. The dashed blue line in panel (c) shows the two-photon reflection density~\cite{SI}. (e) Upper (top curve) and lower bound (bottom curve) for the detuning guaranteeing the single-photon transmission amplitude to be above $\sqrt{0.9}$ as a function of the coupling ratio. (f) Overlap error in the two-photon state (solid line) and the error in the nonlinear phase shift (dashed line) as functions of the effective pulse width $x_\textrm{pulse}$. Here we have set $\gamma\approx0.62$ such that the nonlinear phase shift is $\pi/4$ for collimated photons.}
\label{fig:qubit_results}
\end{figure}

The transmission coefficient for a single-photon can be analytically solved from the Schr\"odinger equation \cite{fang2014one} yielding
\begin{equation} \label{eq:SE_trans_coeff}
S_{21}^\text{\,q}(k)=\frac{2\delta(k)^3}{[i\Gamma+\delta(k)]\left\{2\delta(k)^2-\gamma\Gamma[\Gamma-i\delta(k)] \right\}},
\end{equation}
where $\delta\left(k\right)=vk-\Omega$ is the detuning between the frequencies of the photons and the qubits. In Figs.~\ref{fig:qubit_results}(a,b) we show the absolute value and the phase of the transmission coefficient $S_{21}^\text{\,q}$ as a function of the detuning $\delta$ and the coupling ratio $\gamma$. We observe full single-photon transmission for detuning $\delta_{\textrm{f}}=\pm \sqrt{\gamma/\left(2-\gamma\right)}\Gamma$. The corresponding phase shift for a photon with the wavenumber $k_\textrm{f}=\left[\delta_\mathrm{f}\left(\gamma, \Gamma\right)+\Omega \right]/v$ is given by $\arg\left[S_{21}^\text{\,q}(k_\text{f})\right]=\pm\arctan\left[ \sqrt{2\gamma-\gamma^2}/\left(1-\gamma\right)\right]$. Note that at full transmission, $\delta_\mathrm{f}$ is real-valued only if the coupling ratio obeys $0\leq \gamma < 2$.

In the two-photon case, it is convenient to treat the qubits as bosonic sites by replacing the qubit operators, $\hat{\sigma}_j^+$ and $\hat{\sigma}_j^-$, with bosonic ladder operators, $\hat{d}_j^\dagger$ and $\hat{d}_j$, respectively. Multiple excitations of the sites are prevented by introducing an interaction potential~\cite{longo2010few} $\hat{\mathcal{V}}_j = \frac{U}{2}\hat{d}_j^\dagger\hat{d}_j\left(\hat{d}_j^\dagger\hat{d}_j-1\right)$ for each qubit to the Hamiltonian. The modified Hamiltonian is equivalent to the original one in the limit of infinite interaction strength $U$, which we apply at the end of our calculation. The two-photon eigenstates of the noninteracting  Hamiltonian, i.e., Eq.~(\ref{eq:finalhamiltonian}) with qubits treated as bosonic sites, are direct products of the single-photon states. Utilizing the Lippmann--Schwinger formalism~\cite{zheng2013persistent,fang2014one}, we study how the product state is modified by the interaction potential. We find the transmission density for the two-photon scattered state to be~\cite{SI} $S_{21}^\text{\,q}(k)S_{21}^\text{\,q}(k)+B(k,x_{\mathrm{p}2}-x_{\mathrm{p}1})$, where $x_{\mathrm{p}i}$ is the position of the $i$:th photon, and $B(k,x_{\mathrm{p}2}-x_{\mathrm{p}1})$ is a nonlinear correction to the transmission density. Here, the correction term arises from the effective photon--photon interactions mediated by the qubits: the transmission density for the noninteracting two-photon state is simply $S_{21}^\text{\,q}(k)S_{21}^\text{\,q}(k)$. Note that for interacting photons, the relative amplitude of the transmitted part of the wavefunction cannot be directly interpreted as a transmission coefficient, since the amplitude depends on the spatial separation of the photons. Thus we study the transmission density along the coordinate $x_{\text{p}2}-x_{\text{p}1}$ which may locally exceed unity.

Let us focus on the ideal case of collimated photons, i.e., $x_{\mathrm{p}1}=x_{\mathrm{p}2}$.
In Fig.~\ref{fig:qubit_results}(d), we show the effect of the nonlinear correction to the phase of the scattered two-photon state. We observe that the phase shift is indeed nonlinear, and at $\gamma\approx 0.21$ we obtain the maximum nonlinearity of approximately 3$\pi$/10 radians. The so-called reflection density~\cite{SI} is finite for all $0 \le \gamma < 2$. As $\gamma$ approaches value 2, the reflection density vanishes along with the nonlinearity as shown in Figs.~\ref{fig:qubit_results}(c,d).

In Fig.~\ref{fig:qubit_results}(e), we observe that the single-photon transmission $|S_{21}^\text{\,q}(k)|$ is above 90\% for a very wide range of detunings and coupling ratios.
Another source of error is the dependence of the two-photon output state $\ket{2}_\textrm{out}$ on the spatial separation of the photons. We quantify the resulting overlap error as
\begin{align} \label{eq:qubit_spatial_error}
\mathcal{E}_\text{o} =& 1-\frac{1}{A}\left|\prescript{}{\textrm{out}}{\braket{2|2}_{\textrm{out}}^{\textrm{ideal}}}\right|,
\end{align}
where $\ket{2}_\textrm{out}^\text{ideal}$ is the ideal output state, the normalization coefficient is given by $A=\sqrt{\leftidx{^\textrm{ideal}_\textrm{\phantom{d}out}}{\braket{2|2}}{^\textrm{ideal}_\textrm{out}} \leftidx{_\textrm{out}}{\braket{2|2}}{_\textrm{out}}}$, and the integrations are carried out over an effective pulse width $x_\text{pulse}$. Similarly, we quantify the error in the nonlinear phase shift, $\mathcal{E}_\text{p}$, due to finite pulse width as deviation of the spatially correlated phase shift from that of collimated photons averaged over the pulse width \cite{SI}. The errors are shown in Fig.~\ref{fig:qubit_results}(f) as functions of the spatial width of the photon pulses at the input of the phase shifter. The overlap between the actual and the ideal output states is the greatest for narrow pulses, hence yielding the smallest error. Even though the pulse width is much greater than the wavelength, we may still obtain small error since the coupling strength between the side qubits and the waveguide, $\Gamma$, is much smaller than the frequency of the radiation. The error in the nonlinear phase shift increases with the pulse width as expected since the nonlinearity diminishes as the photon separation is increased.

{\it Discussion.}---In summary, we proposed two different designs of the quantum phase gate for propagating microwave photons. We theoretically showed that both exhibit full single-photon transmission with a tunable phase shift, a property that we experimentally demonstrated for the SQUID-based phase shifter. Furthermore, we showed that the qubit-based shifter is nonlinear. Nonlinearity allows for the realization of the controlled-NOT gate~\cite{SI}, which together with single-qubit gates, obtained using already implemented nontunable beam splitters and tunable linear phase shifters, can be used to implement arbitrary many-qubit gates~\cite{PhysRevLett.93.130502}.

Implementation of a many-qubit quantum processor based on propagating microwave photons remains an attractive challenge in the field of quantum technology. On the short term, implementation of arbitrary tunable single qubit-gates and integrated circuits including sources and detectors on the same chip seems plausible. Extension of the three-SQUID design, used here as a tunable phase shifter, to five or seven SQUIDs may allow in the future an even broader range of phase shifts.

{\it Acknowledgments.}---We wish to thank Matti Partanen for technical assistance and Kazuki Koshino,  Scott Glancy, and Dave Rudman for insightful comments on the manuscript. We acknowledge the financial support from European Research Council under Grants No. 278117 ("SINGLEOUT") and No. 681311 ("QUESS") and Academy of Finland under Grant Nos. 251748, 265675, 276528, 284621, and 305237. We acknowledge the provision of facilities by Aalto University at OtaNano -- Micronova Nanofabrication Centre.
\bibliography{references}
\end{document}